\documentstyle[aps,amsfonts,twocolumn,epsfig]{revtex}

\begin{document}

\title{Entanglement of internal and external angular momenta of a single atom}
\author{Ashok Muthukrishnan and C. R. Stroud, Jr.}
\address{The Institute of Optics, University of Rochester,
Rochester, New York 14627 \\ amuthuk@optics.rochester.edu, Tel: (716) 275-2471,
Fax: (716) 244-4936}
\date{Submitted for publication, October 12, 2001}

\maketitle

%**************************************************************
%
\renewcommand{\thesection}{\arabic{section}}
\renewcommand{\thesubsection}{\thesection.\arabic{subsection}}

\newcommand{\hs}[1]{\hspace{#1 ex}}
\newcommand{\vs}[1]{\vspace{#1 ex}}

\newcommand{\ket}[1]{\mbox{$|#1\rangle$}}
\newcommand{\bra}[1]{\mbox{$\langle #1|$}}
\newcommand{\braket}[2]{\mbox{$\langle #1|#2 \rangle$}}
\newcommand{\ketbra}[2]{\mbox{$|#1\rangle\langle #2|$}}
\newcommand{\matrixelem}[3]{\mbox{$\langle #1|#2|#3 \rangle$}}

\newcommand{\cE}{\mbox{${\cal E}$}}
\newcommand{\cO}{\mbox{${\cal O}$}}

\newcommand{\nbar}{\bar{n}}

\newcommand{\Sch}{Schr\"odinger }

\newcommand{\postscript}[1] {\centerline{\epsfbox{#1}}}
%
%******************************************************************************

\widetext \hskip.5in
\begin{minipage}{5.75in}
\begin{abstract}
\vs{-8}

\hs{1.5} We consider the exchange of spin and orbital angular momenta between a
circularly polarized Laguerre-Gaussian beam of light and a single atom trapped
in a two-dimensional harmonic potential. The radiation field is treated
classically but the atomic center-of-mass motion is quantized. The spin and
orbital angular momenta of the field are individually conserved upon
absorption, and this results in the entanglement of the internal and external
degrees of freedom of the atom. We suggest applications of this entanglement in
quantum information processing.\vs{1}

PACS numbers: 42.50.Vk, 32.80.Lg

\end{abstract}
\end{minipage}

\narrowtext \vs{0.8}

%***************************************************************************
\section{Introduction}

The Laguerre-Gaussian (LG) laser modes are known to possess well-defined,
discrete values of orbital angular momentum per unit energy \cite{Allen92}. The
orbital angular momentum of the field is distinct from the spin angular
momentum associated with the polarization state of the field. In the paraxial
limit, the orbital angular momentum is polarization-independent
\cite{Barnett94} and arises solely from the azimuthal phase dependence of the
field mode which gives rise to helical wavefronts.

The interaction of LG modes with atoms has been studied extensively in the
classical limit of the atom as a point particle \cite{Allen96}. It has been
shown that the atom experiences a torque from the radiation pressure force,
which transfers the angular momentum from the laser beam to the atom. This
effect has been indirectly observed in the nonlinear four-wave mixing of LG
modes in a cold atomic sample \cite{Tabosa99}. There have also been proposals
to use LG modes to create vortices in Bose-Einstein condensates
\cite{Marzlin97,Bolda98}, where the orbital angular momentum is transferred
from the laser beam to the vortex trap state.

In this paper we consider the interaction of a circularly polarized LG mode
with a single trapped atom whose center-of-mass (CM) motion is quantized. We
take the trapping potential to be harmonic in two dimensions, which may be
approximated by an optical dipole trap created by a doughnut-shaped LG beam
\cite{Kuga97}. In the paraxial limit, we know that the angular momentum of the
atom-plus-field system is separately conserved for spin and orbital components
\cite{vanEnk94}. We suggest this mechanism as a means for entangling the
internal and external angular momenta of the atom.

In section~\ref{sec-basis}, we introduce the basis states for the
center-of-mass and electronic degrees of freedom of the atom. In
section~\ref{sec-hamiltonian} we derive the Hamiltonian that describes the
interaction of the atom with a circularly polarized LG beam. The quantization
of the atomic position is made in the limit that the size of the center-of-mass
wave function is small compared to the beam waist. In
section~\ref{sec-entanglement} we show how the interaction leads to an
\linebreak

\begin{minipage}{58ex}\vs{24}
\end{minipage}

\noindent entanglement between the internal and external states of the atom.
Finally in section~\ref{sec-discussion}, we consider the relevance of this
phenomenon for quantum information applications.

%***************************************************************************
\section{Basis states}\label{sec-basis}

The Hamiltonian for a harmonic trapping potential in two dimensions is
\begin{equation}\label{HamiltonianCM}
    \hat{H}_{\mathrm{CM}}
        = \frac{1}{2m} (\hat{P}_X^2 + \hat{P}_Y^2) +
          \frac{1}{2} m \nu^2 (\hat{X}^2 + \hat{Y}^2),
\end{equation}
where $m$ is the mass of the atom and $\nu$ is the radial trap frequency. We
assume that the atom is tightly confined along the trap axis, $\nu_z \gg \nu$.
The Hamiltonian in Eq.~(\ref{HamiltonianCM}) describes two independent
one-dimensional harmonic oscillators along the Cartesian axes with annihilation
operators
\begin{eqnarray}\label{annihilationX}
    \hat{a}_X =
        \frac{1}{\sqrt{2}} \left(\frac{\hat{X}}{R_0}
        + i \frac{\hat{P}_X}{R_0/\hbar} \right),
\\ \label{annihilationY}
    \hat{a}_Y =
        \frac{1}{\sqrt{2}} \left(\frac{\hat{Y}}{R_0}
        + i \frac{\hat{P}_Y}{R_0/\hbar} \right),
\end{eqnarray}
where $R_0 = \sqrt{\hbar/m\nu}$ sets the scale for the radial size of the trap.
Since we are interested in the angular momentum of the atom as a whole, it is
convenient to introduce the operators
\begin{equation}\label{annihilation+-}
    \hat{a}_{\pm} =
        \frac{1}{\sqrt{2}} (\hat{a}_X \mp i\hat{a}_Y),
\end{equation}
which serve to raise ($\hat{a}_+^{\dagger}$ and $\hat{a}_-$) and lower
($\hat{a}_-^{\dagger}$ and $\hat{a}_+$) the angular momentum $L_z$ along the
trap axis. We can show that
\begin{equation}\label{AngularmomentumExt}
    \hat{L}_Z = \hbar (\hat{a}_+^{\dagger} \hat{a}_+
                - \hat{a}_-^{\dagger} \hat{a}_-),
\end{equation}

% ****************************************************
%
\begin{figure}[htp]
\postscript{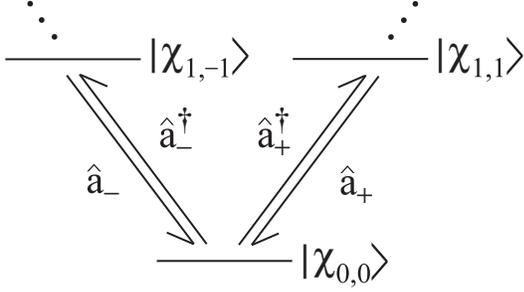}
\caption{Ground and first excited states of the trap} \vs{2}
\label{figure1}
\end{figure}
%
% ****************************************************

\noindent where $n_+ = \langle \hat{a}_+^{\dagger} \hat{a}_+ \rangle$ and $n_-
= \langle \hat{a}_-^{\dagger} \hat{a}_- \rangle$ count the number of right and
left circular quanta respectively. The Hamiltonian for the two-dimensional
oscillator in terms of these operators is
\begin{equation}\label{HamiltonianCM2}
    \hat{H}_{\mathrm{CM}}
        = \hbar \nu (\hat{a}_+^{\dagger} \hat{a}_+
          + \hat{a}_-^{\dagger} \hat{a}_-
          + 1).
\end{equation}
The center-of-mass eigenstates of the atom can be written in terms of the
energy and angular momentum quantum numbers, $N = n_+ + n_-$ and $M = n_+ -
n_-$.  For a fixed value of $N \ge 0$, there are $N+1$ degenerate angular
momentum states for which $M = -N, -N+2, \ldots, N$. The ground state and the
first excited states of the trap in polar coordinates, $R = \sqrt{X^2 + Y^2}$
and $\Phi = \tan^{-1} (Y/X)$, are given by the wave functions
\begin{eqnarray}\label{trapstate0}
    \chi_{0,0}(R,\Phi) & = & \frac{1}{R_0 \sqrt \pi} \mbox{ } \exp\left(-\frac{R^2}{2
    R_0^2}\right),
\\ \label{trapstate1}
    \chi_{1,\pm 1}(R,\Phi) & = & \frac{1}{R_0 \sqrt \pi}
    \left(\frac{R}{R_0}\right) \exp\left(-\frac{R^2}{2 R_0^2} \pm i\Phi\right).
\end{eqnarray}
The energy levels for these states and the corresponding transition operators
are shown in figure~\ref{figure1}. The amplitudes of $\chi_{1,\pm 1}$ are
shaped in the form of a doughnut with a null at the center, and the azimuthal
phase is determined by the angular momentum $M=\pm 1$. In general, the
$\chi_{N,M}$ wave functions are given by Laguerre-Gaussian modes
\cite{Wallace72}.

To describe the internal angular momentum of the atom, we need to introduce a
basis of electronic states. In this paper we consider the hydrogenic circular
states \cite{Hulet83}, which have the maximum angular momentum component
$m\hbar$ along the $z$ axis for a given principal quantum number $n=m+1$. Only
neighboring circular states are coupled according to dipole selection rules and
hence these states serve as good approximations to two-level systems. The
internal angular momentum in this basis is given by
\begin{equation}\label{AngularmomentumInt}
    \hat{l}_z = \sum_{m=0}^{\infty} m\hbar \mbox{ } \ketbra{m}{m}.
\end{equation}
For $\Delta m = \pm 1$ transitions, the dipole moment of the atom in the
circular-state basis can be written as

% ****************************************************
%
\begin{figure}[htp]
\postscript{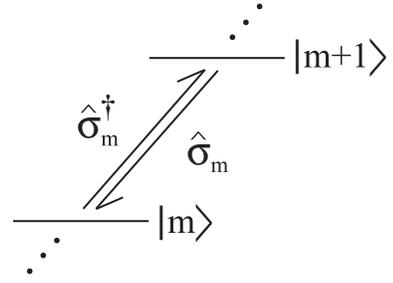}
\vs{1.4}\caption{Atomic circular states \ket{m} = \ket{n=m+1,m,m}} \vs{2}
\label{figure2}
\end{figure}
%
% ****************************************************

\begin{equation}\label{dipolemoment}
    \hat{\bf d}_m = \frac{1}{2} \sum_{m} d_m
    [({\bf x} + i {\bf y}) \hat{\sigma}_{m} + ({\bf x} - i {\bf y})
    \hat{\sigma}_{m}^{\dagger}],
\end{equation}
where $d_m = e (x-iy)_{m,m+1} = e(x+iy)_{m+1,m}$ is the dipole moment matrix
element, and $\hat{\sigma}_m = \ketbra{m}{m+1}$ is the lowering operator for
the transition between neighboring circular states. As figure~\ref{figure2}
shows, the effect of $\hat{\sigma}_m^{\dagger}$ and $\hat{\sigma}_m$ is to
raise and lower the internal angular momentum of the atom in a given circular
state.

%***************************************************************************
\section{Interaction Hamiltonian}\label{sec-hamiltonian}

The laser field is taken to be in an LG mode $(l,p)$. These modes are
characterized by an angular momentum $l\hbar$ about the propagation axis, and
have $p+1$ radial nodes in the transverse intensity distribution
\cite{Allen92}. For simplicity, we consider the $p=0$ case, which corresponds
to a doughnut-shaped intensity distribution for $l \ne 0$. The transverse
profile of the mode at the beam waist $w_0$ is given by
\begin{equation}\label{LGmode}
    u_l(R,\Phi) = \cE_l \left(\frac{R}{w_0}\right)^{|l|}
    \exp\left(-\frac{R^2}{w_0^2} + i l\Phi \right).
\end{equation}
We are interested in the limit in which the size of the trapped atom $R_0$ is
small compared to the radius of the LG mode. This affords a linearization of
the interaction Hamiltonian analogous to the Lamb-Dicke limit in trapping and
cooling. In this limit, we are justified in expanding the LG mode in powers of
$R/w_0$,
\begin{equation}\label{LGmodeexpansion}
    u_l(R,\Phi) = \cE_l \left(\frac{R}{w_0}\right)^{|l|} \exp(il\Phi)
    + \cO\left[(R/w_0)^{|l|+2}\right].
\end{equation}
Keeping only the leading-order term in this expansion, we quantize the atomic
center-of-mass position as follows. For $l = \pm |l|$,
\begin{equation}\label{quantization}
    \left(\frac{\hat{R}}{w_0}\right) \exp(\pm i\hat{\Phi})
    = \frac{\hat{X} \pm i\hat{Y}}{w_0}
    = \eta (\hat{a}_{\pm}^{\dagger} + \hat{a}_{\mp}),
\end{equation}
where $\eta = R_0/w_0$ compares the ground-state trap size to the beam waist.
When $\eta=0$, the atom behaves like a point particle. When $\eta \ll 1$, we
can treat the interaction Hamiltonian to lowest order in the center-of-mass
position operators for the atom.

To couple neighboring circular states in the atom, we need a left-circularly
polarized field. Using the truncated form of the LG mode $u_l$ in
Eq.~(\ref{LGmodeexpansion}), we write the electric field on the trap plane for
$l=\pm|l|$ as
\begin{equation}\label{field}
     {\bf E}_l = \cE_l \hs{0.2} \eta^{|l|}
    (\hat{a}_{\pm}^{\dagger} + \hat{a}_{\mp})^{|l|}
    ({\bf x} + i {\bf y}) \exp(-i\omega t) + \mathrm{h.c.},
\end{equation}
where h.c. denotes hermitian conjugate. The coupling between the LG mode and
the trapped atom is described by the ${\bf d}\cdot{\bf E}$ Hamiltonian. Using
the dipole moment and the field vector from Eq.~(\ref{dipolemoment}) and
Eq.~(\ref{field}), we find that for $l=\pm |l|$, the interaction Hamiltonian is
given by
\begin{eqnarray}
    \hat{H}_{\mathrm{int}} & = & -\hat{\bf d} \cdot {\bf E}_l
\nonumber \\*
    & = & - \frac{\hbar}{2} \sum_m \eta^{|l|} \Omega_{m,l}
    (\hat{a}_{\pm}^{\dagger} + \hat{a}_{\mp})^{|l|} \hat{\sigma}_m^{\dagger}
    \exp(-i\omega t) \hs{0.5}+\hs{0.5}\mathrm{h.c.},
\nonumber \\*[-1ex] \label{Hint}
\end{eqnarray}
where $\Omega_{m,l} = 2 d_m \cE_l/\hbar$ is the Rabi frequency, and we have
used the vector identities $({\bf x} \pm {\bf y}) \cdot ({\bf x} \pm {\bf y}) =
0$ and $({\bf x} \pm {\bf y}) \cdot ({\bf x} \mp {\bf y}) = 2$.

We now specialize to the case of a two-level system formed by two neighboring
circular states $m$ and $m+1$. In the interaction picture, the states evolve
only according to $H_{\mathrm{int}}$. The atomic operators evolve as
$\hat{\sigma}_m(t) = \hat{\sigma}_m \exp[-i(\omega_{m+1} \hs{-0.2}-\hs{-0.2}
\omega_m)t]$, where $\omega_m$ are the atomic frequencies of the states.
Similarly, the center-of-mass operators evolve as $\hat{a}_{\pm}(t) =
\hat{a}_{\pm} \exp(-i\nu t)$, where $\nu$ is the trap frequency. Consider the
situation where the field is tuned to the $|l|^{\mathrm{th}}$ sideband below
the atomic resonance,
\begin{equation}\label{frequency}
    \omega = (\omega_{m+1} \hs{-0.2}-\hs{-0.1} \omega_m) - |l|\nu.
\end{equation}
In the rotating-wave approximation, we ignore counter-rotating terms in the
interaction Hamiltonian and are left with only the two circular states $m$ and
$m+1$ contributing to the sum in Eq.~(\ref{Hint}). Furthermore, if the field is
sufficiently narrow in spectrum compared to the trap frequency, only the
$|l|^{\mathrm{th}}$ power of the operators $\hat{a}_{\pm}$ contribute to the
interaction in Eq.~(\ref{Hint}), assuming $\eta^{|l|} \Omega_{m,l} \ll \nu$,
and we can ignore the cross terms. The interaction Hamiltonian simplifies to
\begin{equation}\label{Hint2}
    \hat{H}_{\mathrm{int}} = - \frac{\hbar}{2} \mbox{ } \eta^{|l|} \Omega_{m,l}
    \mbox{ } \hat{a}_{\mp}^{|l|} \hat{\sigma}_m^{\dagger}
    + \mbox{ } \mathrm{h.c.}.
\end{equation}
To interpret the interaction in physical terms, recall that
$\hat{a}_{+}^{\dagger}$ and $\hat{a}_{-}$ raise the center-of-mass angular
momentum, while $\hat{a}_{-}^{\dagger}$ and $\hat{a}_{+}$ lower it. This can be
seen from Eq.~(\ref{AngularmomentumExt}). Similarly, $\hat{\sigma}_m^{\dagger}$
($\hat{\sigma}_m$) raises (lowers) the internal angular momentum of the atom in
the two circular states.  Thus the Hamiltonian in Eq.~(\ref{Hint2}) clearly
shows that the orbital angular momentum of the LG mode is transferred to the
external angular momentum of the atom, while the spin angular momentum
associated with circular polarization is transferred to the internal angular
momentum of the atom. The spin and orbital components are separately conserved
in the paraxial limit.

Choosing the orbital angular momentum of the LG mode to be positive or
negative, $l=\pm |l|$, correlates the change in the internal and external
angular momenta of the atom, $\Delta m$ and $\Delta M$. For example, when $l<0$
the external angular momentum is raised whenever the internal angular momentum
is lowered, and vice versa. The choice of field frequency in
Eq.~(\ref{frequency}) corresponds to tuning to the $|l|^{\mathrm{th}}$ sideband
on the lower side of the atomic resonance. This choice governs the parity of
the transitions between the center-of-mass states as shown in
figure~\ref{figure1}, and correlates the change in the energy and angular
momentum of the trap, $\Delta N$ and $\Delta M$.

%***************************************************************************
\section{Entanglement}\label{sec-entanglement}

We use Eq.~(\ref{Hint2}) as the starting point for a discussion of quantum
entanglement between the internal and external angular momenta of the atom in
the trap. Consider $l=-1$, which gives the left-circularly polarized LG field a
net angular momentum of zero. In this case, the change in internal and external
angular momenta of the atom are equal in magnitude but opposite in sign. The
time evolution operator in this case is given by
\begin{eqnarray}
    \hat{U}_{\mathrm{int}}(t) & = & \exp(-i\hat{H}_{\mathrm{int}} t/\hbar)
\nonumber \\ \label{Uint}
    & = & \exp[i \frac{\eta t}{2} (\Omega_{m,-1} \hat{a}_+ \hat{\sigma}_m^{\dagger}
    + \Omega_{m,-1}^* \hat{a}_+^{\dagger} \hat{\sigma}_m)].
\end{eqnarray}
Consider the action of this operator on the state $\ket{m}\ket{\chi_{0,0}}$.
Since the internal angular momentum of the atom can only be raised by
$\hat{\sigma}_m^{\dagger}$, the external angular momentum has to be lowered by
$\hat{a}_+$. However $\ket{\chi_{0,0}}$ is the lowest energy state of the trap
and cannot be further reduced in energy. Hence the state
$\ket{m}\ket{\chi_{0,0}}$ does not evolve in time according to this
interaction. This restriction does not apply to the state
$\ket{m+1}\ket{\chi_{0,0}}$, since the atom is in the higher angular momentum
circular state to begin with, and we find Rabi oscillations between states
$\ket{m+1}\ket{\chi_{0,0}}$ and $\ket{m}\ket{\chi_{1,1}}$. To summarize, we
find that
\begin{eqnarray}\label{evolution1}
    \hat{U}_{\mathrm{int}}(t) \hs{0.2}\ket{m}\ket{\chi_{0,0}} & = &
    \ket{m}\ket{\chi_{0,0}},
\\*[1ex]
    \hat{U}_{\mathrm{int}}(t) \hs{0.2}\ket{m+1}\ket{\chi_{0,0}} & = &
    \cos(\eta \Omega t/2) \hs{0.3} \ket{m+1}\ket{\chi_{0,0}}
\nonumber \\* \label{evolution2}
    & & \hs{0.3}+\hs{0.7}
    i e^{i\phi} \sin(\eta \Omega t/2) \hs{0.3} \ket{m}\ket{\chi_{1,1}},
\end{eqnarray}
where we have defined $\Omega_{m,-1} = \Omega e^{i\phi}$.
Equations~(\ref{evolution1}) and (\ref{evolution2}) give the basic ingredients
for quantum control of the selected internal and external states of the atom.
When the trap is in the ground state and the atom is prepared in a coherent
superposition of the circular states $m$ and $m+1$, a $\pi$-pulse transfers
this coherence to the center-of-mass state of the atom in the trap,
\begin{eqnarray}
    & \hat{U}_{\mathrm{int},\pi} \hs{-0.4}: \hs{0.8}
     [c_m \ket{m} + c_{m+1}\ket{m+1}] \mbox{ }\ket{\chi_{0,0}} &
\nonumber \\* \label{pipulse}
    & \hs{10} \mapsto
    \ket{m}\mbox{ } [c_m \ket{\chi_{0,0}} + c_{m+1}\ket{\chi_{1,1}}], &
\end{eqnarray}
where we have chosen $\phi = -\pi/2$. Alternatively, if the atom is in the
upper circular state $m+1$ and the trap is in the ground state, a $\pi/2$-pulse
creates maximal entanglement between the internal and external states,
\begin{equation}\label{pi/2pulse}
    \hat{U}_{\mathrm{int},\pi/2} \hs{-0.3}: \ket{m+1} \ket{\chi_{0,0}}
    \mapsto
    \frac{1}{\sqrt{2}} \left[\ket{m+1} \ket{\chi_{0,0}}
        +  \ket{m} \ket{\chi_{1,1}} \right] \hs{-0.2},
\end{equation}
where $\phi = -\pi/2$ again. We have to be in the adiabatic limit where the
pulse length is long enough that the spectrum does not overlap neighboring trap
states in energy.

Equation~(\ref{pi/2pulse}) is the main result of this paper, that we can in
principle generate states of a single atom that are entangled in internal and
external angular momenta using a circularly polarized LG mode. This is a new
form of entanglement that relies on the conservation of angular momentum rather
than energy. The two observables that are entangled are $L_z$ and $l_z$,
defined in Eqs.~(\ref{AngularmomentumExt}) and (\ref{AngularmomentumInt})
respectively.

The experimental difficulty is in measuring the quantized center-of-mass state
of the atom in the trap. A direct observation of the trap state may be
engineered as follows. When the atoms are released from an excited trap state
\ket{\chi_{1,1}}, they escape with a net linear momentum in the azimuthal
direction, which may be detected by time-of-flight measurements using a
suitably positioned detector array. An indirect observation of the entanglement
present in Eq.~(\ref{pi/2pulse}) is possible using a weak probe pulse resonant
with the circular states $m+1$ and $m+2$. In this case, only the state
$\ket{m+1}\ket{\chi_{1,1}}$ is affected by the pulse, and the absorption of a
photon would distinguish this state from $\ket{m+1}\ket{\chi_{0,0}}$.

%***************************************************************************
\section{Discussion}\label{sec-discussion}

It is intriguing to consider the application of the ideas in this paper to
quantum information processing. Rydberg circular states are extremely
long-lived, with radiative lifetimes of the order of $10$ milliseconds even for
$n=30$, increasing as $n^5$ for larger $n$. The two circular states $m$ and
$m+1$ may be thought of as a qubit, and the interaction with the LG mode
provides a controlled coupling to the second qubit formed by the ground and
first excited state of the trap. In this context, Eqs.~(\ref{evolution1}) and
(\ref{evolution2}) allow arbitrary controlled unitary operations, where the
internal state of the atom plays the role of the control qubit.

One possibility to scale up this scenario is to consider two or more atoms
individually trapped and manipulated in this manner. A coupling between two
atoms may be achieved by entangled photons in LG modes, as demonstrated
recently in the parametric downconversion experiment of Ref.~\cite{Mair01}. The
trap states of each atom can play the role of an auxiliary or intermediary
qubit that enables information processing in the internal states of the atom.
Decoherence issues involved with trapping and cooling the atom to the
center-of-mass ground state benefit from the weak coupling of neutral atoms to
the environment.

Lastly, we highlight the benefits of going beyond two internal states (beyond
qubits) in the atom. We chose the circular states because they made a good
two-level system. However, there are $n^2$ angular momentum states in the atom
for each principal quantum number $n$, all of which are degenerate in hydrogen.
This allows for the possibility of simultaneous control of these states and
entanglement with the trap states. Angular momentum entanglement is
particularly suited for large-scale information processing in the atom.

\section*{Acknowledgements}

This work was supported by the Army Research Office through the MURI Center for
Quantum Information.

\vs{-0.5}
%******************************************************************************
\makeatletter

\end{document}